\documentclass[aps,pre,preprint,showkeys,showpacs,preprintnumbers,superscriptaddress,amsmath,amssymb]{revtex4-1}
\usepackage{graphicx}
\usepackage{dcolumn}
\usepackage{bm}

\usepackage{color}
\usepackage[normalem]{ulem}
\usepackage{}

\begin{document}
\title{Suppression of macroscopic oscillations in mixed populations of active and inactive oscillators coupled through lattice Laplacian}
\author{Ikuhiro Yamaguchi}
\email[]{ikuhiro@p.u-tokyo.ac.jp}
\affiliation{
Graduate School of Education, The University of Tokyo, 7-3-1 Hongo, Bunkyo-ku, Tokyo 113-0033, Japan
}

\author{Takuya Isomura}
\affiliation{
Graduate School of Frontier Science, 
The University of Tokyo, 5--1--5 Kashiwanoha, Kashiwashi, Chiba 277--8563,
Japan
}

\author{Hiroya Nakao}
\affiliation{Graduate School of Information Science and Engineering, Tokyo Institute of Technology, Tokyo 152-8522, Japan
}

\author{Yutaro Ogawa}
\affiliation{Graduate School of Frontier Science, 
The University of Tokyo, 5--1--5 Kashiwanoha, Kashiwashi, Chiba 277--8563,
Japan
}

\author{Yasuhiko Jimbo}
\affiliation{Graduate School of Engineering, The University of Tokyo, 7-3-1, Hongo, Bunkyo-ku, Tokyo 113-8656, Japan}

\author{Kiyoshi Kotani}
\affiliation{Research Center for Advanced Science and Technology, The University of Tokyo, 4-6-1, Komaba, Meguro-ku, Tokyo 153-8904, Japan}
\affiliation{PRESTO, Japan Science and Technology Agency, 4-1-8, Honcho, Kawaguchi-shi, Saitama 332-0012, Japan}

\date{\today}

\begin{abstract}
We consider suppression of macroscopic synchronized oscillations in mixed populations of active and inactive oscillators with local
diffusive coupling, described by a lattice complex Ginzburg-Landau model with discrete Laplacian in general dimensions. Approximate expression for the stability of the non-oscillatory stationary state is derived on the basis of the generalized free energy of the system. We show that an effective wavenumber of the system determined by the spatial arrangement of the active and inactive oscillators is an decisive factor in the suppression, in addition to the ratio of active population to inactive population and relative intensity of each population. The effectiveness of the proposed theory is illustrated with a cortico-thalamic model of epileptic seizures, where active and inactive oscillators correspond to epileptic foci and healthy cerebral cortex tissue, respectively.
\end{abstract}


\maketitle

\section{Introduction}

Coupled-oscillator models have been widely used in the studies of a variety of rhythmic phenomena in physics, chemistry, and biology~\cite{kuramoto2012chemical, winfree2001geometry,ashwin2016mathematical}.
While many results for coupled oscillators have been obtained for phase models, in which the amplitude variable of oscillators is eliminated and only the phase of oscillators is considered, the effect of amplitude degrees of freedom often leads to intriguing collective dynamics that cannot occur in phase models.
            
Regarding collective oscillations in coupled oscillators with amplitude degrees of freedom, Daido and Nakanishi performed a pioneering study that revealed the effect of deteriorated inactive oscillators on macroscopic collective oscillation of the whole system, which they called aging transition~\cite{daido2004aging}. In their model, macroscopic synchronized oscillation turns into a quiescent state as the proportion of inactive elements exceeds a certain critical value when coupling strength is sufficiently large. 

The aging transition framework of Daido and Nakanishi, originally analyzed for globally coupled oscillators~\cite{daido2004aging, pazo2006universal, tanaka2010phase, daido2011strong, daido2013onset}, has been generalized to various kinds of coupled-oscillator models with different coupling topologies. For example, aging transitions in a one-dimensional ring of locally coupled oscillators~\cite{daido2008aging} and in networks of coupled oscillators with random coupling~\cite{morino2011robustness,tanaka2012dynamical} have been analyzed. 
However, compared with global coupling, mathematical treatment of other coupling topologies is difficult and relied mostly upon numerical simulations in most cases. As for coupled oscillators with lattice Laplacian coupling, which is a discretized version of ordinary diffusion in continuous media, spatial arrangement of the active and inactive oscillators decisively affects the aging transition in contrast to the case of global coupling.

The main aim of this paper is providing a theory to deal with heterogeneous spatial arrangement of active and inactive oscillators. We consider the suppression of macroscopic oscillations in mixed populations of active and inactive oscillators (i.e., aging transition) coupled through lattice Laplacian in general dimensions. By introducing a generalized free energy (Lyapunov function) of the system and approximating it with two variables, we show that an effective wavenumber of the system can be introduced, which characterizes the arrangement of the active and inactive oscillators, and can be used for determining whether suppression of macroscopic oscillation occurs or not.
 
Another aim of this paper is to show how the aging transition framework with the proposed theory provides insights into realistic situations in a cortico-thalamic model of epileptic seizures.
Though the oscillating state and quiescent state were interpreted as alive and dead respectively in the original study~\cite{daido2004aging}, the aging transition framework can also be interpreted in the opposite sense.
That is, we consider whether pathological macroscopic oscillation caused by pathological oscillators can be suppressed when the proportion of the healthy oscillators is increased~\cite{kim2009dynamics}.
Using the proposed theory, we analyze a cortico-thalamic model of epileptic seizures, where active and inactive oscillators correspond to epileptic foci and healthy region, respectively.

\section{Theory}
\subsection{Model}

We consider a mixed population of active and inactive oscillators, where the active group is denoted as S (seizure) and the inactive group is denoted as H (healthy).
The oscillators are placed on a $p$-dimensional hypercubic lattice of $N^{p}$ sites with a lattice constant $d$, where $L=Nd$ is the side length of the hypercube, and periodic boundary conditions are assumed.

The system is described by the following lattice complex Ginzburg-Landau model on a $p$-dimensional lattice ($p$ is a natural number) of lattice constant $d$:
\begin{eqnarray}
\frac{dB}{dt}=-i \Omega B + (a+ib) \left[ \left( \frac{1}{2} \mu+\frac{3}{8} \epsilon \left | B \right |^{2} \right ) B + \frac{1}{2} r_{e}^{2} \Delta B \right ].
\label{CGL_0}
\end{eqnarray}
Here, $B=B(\bm{j}; t)$ is the complex amplitude of the oscillator at lattice coordinate $\bm{j}=(j_1, j_2, \dots, j_p)$ and at time $t$, $\Omega$ is the natural frequency of the oscillator at the onset of oscillation, $a >0$, $b$ and $\epsilon <0$ are real parameters, and $r_{e}$ represents a diffusion constant.
The Hopf bifurcation parameter $\mu  = \mu({\bm j})$ determines if the oscillator is active ($\mu>0$) or inactive ($\mu<0$), and $\Delta$ denotes the $p$-dimensional discrete (lattice) Laplacian defined as
\begin{equation}
\Delta B(\bm{j}) = \frac{1}{d^2} \sum_{q=1}^{p}\{ B(j_1, \dots, j_q-1, \dots, j_p) + B(j_1, \dots, j_q+1, \dots, j_p) - 2 B(\bm{j}) \}.
\end{equation}

For simplicity, we restrict our investigation to the case where all oscillators have the same frequency parameter $\Omega$.
Then, by introducing a rotating frame with frequency $\Omega$ and denoting $B=Ae^{i \Omega}$, we obtain
\begin{eqnarray}
\frac{dA}{dt}= (a+ib) \left[
 \left( \frac{1}{2} \mu+\frac{3}{8} \epsilon \left | A \right |^{2} \right ) A + \frac{1}{2} r_{e}^{2} \Delta A \right ].
\label{CGL}
\end{eqnarray}
We also assume that the characteristics of the oscillators are binary. That is,
if the oscillator belongs to group S, then its bifurcation parameter satisfies $\mu=\mu_{S}>0$; if the oscillator belongs to group H, $\mu=\mu_{H}<0$. All parameters other than $\mu$ are identical in both groups.

Depending on the parameters and the spatial arrangement of the active and inactive oscillators, the system is in the quiescent state ($A=0$) or in the active state ($|A|>0$).
When the system is in the active state, that is, when the oscillators undergo collective oscillations, we can assume complete synchronization of the oscillators in each group because the frequency parameters of the oscillators are the same and the diffusion constant is real.
That is, we suppose that the oscillation amplitudes of the oscillators are uniform in each group, i.e., $A=A_{S} e^{i \theta (t)}$ in group S and $A=A_{H} e^{i \theta (t)}$  in group H, where $A_S$ and $A_H$ are real amplitudes and the real-valued function $\theta(t)$ represents the phase of the oscillation in the rotating frame. Namely, we assume that the oscillators are approximately binarized into either of the two groups.
This binarization assumption holds strictly when $r_{e}$ is infinitesimal or when the oscillators of group S and H are arranged in the smallest possible checkered pattern.
Though some error may generally be expected, this assumption also works reasonably well for other arrangements and makes the problem analytically tractable.

\subsection{Two-variable free energy}

It is known that Eq.~(\ref{CGL}) has a Lyapunov function, or the generalized free energy (GFE), which monotonously decreases with the evolution of the system~\cite{aranson2002world}. 
As we derive below, we can approximate the GFE of Eq.~(\ref{CGL}) by a two-variable function $F(A_{S}, A_{H})$ as
\begin{eqnarray}
\frac{4F}{L^{p}}=-(\alpha_{S} \mu_{S} A_{S}^{2}+\alpha_{H} \mu_{H} A_{H}^{2})+\alpha_{S} \alpha_{H} r_{e}^{2} k_{m}^{2} (A_{S}-A_{H})^2 + 
O(|A|^{4}),
\label{FE}
\end{eqnarray}
under the \textit{two-group approximation} that we introduce below, where $O(|A|^4)$ represents $4$th-order terms in $A_{S}$ and $A_{H}$.
Here, $\alpha_{H}(>0)$ and $\alpha_{S}(>0)$ are the proportions of the oscillators in group S and group H, respectively ($\alpha_{H}+\alpha_{S}=1$), and the parameter $k_m$ represents the \textit{effective wavenumber} characterizing the spatial arrangement of group S and H defined below.

The approximate GFE Eq.~(\ref{FE}) can be obtained as follows. The exact GFE of Eq.~(\ref{CGL}) is given by ~\cite{aranson2002world}
\begin{eqnarray}
F &=& d^{p} \sum_{\bm{j}} f(A(\bm{j})), \cr
f(A) &=& \frac{1}{2}\left[-\frac{1}{2}\mu \left | A \right |^{2} -\frac{3}{16} \epsilon \left | A \right |^{4} + \frac{1}{2} r_{e}^{2} \left | \nabla A \right|^{2} \right] 
\label{CE_F},
\end{eqnarray}
where $\nabla$ denotes the discrete gradient operator on the $p$-dimensional lattice defined as
\begin{equation}
\nabla A(\bm{j}) = \frac{1}{d} \left( A(j_1+1, j_2, \dots, j_p) - A(\bm{j}),\dots, A(j_1, \dots, j_p+1) - A(\bm{j}) \right).
\label{DGra}
\end{equation}
Using the above GFE, Eq.~(\ref{CGL}) and its complex conjugate can be expressed as
\begin{eqnarray}
\frac{dA}{dt}=-(a+ib) \frac{\partial F}{\partial A^{*}}, \; \; \frac{dA^{*}}{dt}=-(a-ib) \frac{\partial F}{\partial A},
\end{eqnarray}
where $A = A(\bm{j})$ and $A^{*} = A^{*}(\bm{j})$.
Note that $F$ is monotonically decreasing:
\begin{eqnarray}
\begin{aligned}
\frac{dF}{dt}&=d^{p} \sum_{\bm{j}} \left ( \frac{\partial f}{\partial A}\frac{dA}{dt}+\frac{\partial f}{\partial A^{*}}\frac{dA^{*}}{dt} \right ) \\
&=-d^{p} \sum_{\bm{j}} \left ( \frac{\partial f}{\partial A} (a+ib) \frac{\partial f}{\partial A^{*}} +\frac{\partial f}{\partial A^{*}} (a-ib) \frac{\partial f}{\partial A} \right ) \\
&=-2ad^{p}\sum_{\bm{j}} \left | \frac{\partial F}{\partial A} \right |^{2} \leq 0,
\end{aligned}
\end{eqnarray}
where $a>0$ is used.
Using the binarization assumption, the summation of the square amplitude in Eq.~(\ref{CE_F}) can be approximated as
\begin{eqnarray}
d^{p} \sum_{\bm{j}}\mu |A|^{2} &=& d^{p} \sum_{\bm{j}} \mu(\bm{j})|A(\bm{j})|^{2} \nonumber \\
&\simeq&L^{p} \left( \alpha_{S} \mu_{S}A_{S}^{2}+\alpha_{H} \mu_{H} A_{H}^{2} \right ),
\label{CE_Am}
\end{eqnarray}
which we call the two-group approximation.

To rewrite the summation of the discrete Laplacian term in Eq.~(\ref{CE_F}), we use the discrete Fourier transformation (DFT),
\begin{eqnarray}
\begin{aligned}
A(\bm{j})&=\sum_{\bm{k}} A(\bm{k})e^{i \bm{k}\bm{x}}, \\
\mu (\bm{j})&=\sum_{\bm{k}} \mu (\bm{k})e^{i \bm{k}\bm{x}}, \\
\bm{k}\bm{x} &\equiv \sum_{q=1}^{p} k_{q}j_{q}d.
\end{aligned}
\end{eqnarray}
We can then rewrite the summation in Eq.~(\ref{CE_F}) as
\begin{eqnarray}
\begin{aligned}
&d^{p} \sum_{{\bm j}} |\nabla A|^{2} = d^{p} \left| \nabla \sum_{\bm{k}} A(\bm{k})e^{i \bm{k}\bm{x}} \right |^{2} \\
&=d^{p} \sum_{\bm{j}} \left ( \nabla \sum_{\bm{k}} A(\bm{k}) e^{i \bm{k}\bm{x}}  \right ) \left ( \nabla \sum_{\bm{k'}} A^{*}(\bm{k'})e^{-i \bm{k'}\bm{x}}  \right ) \\
&=d^{p} \sum_{\bm{j}} \left ( \sum_{\bm{k}} i \bm{k}_{\bm{d}} A(\bm{k}) e^{i \bm{k}\bm{x}}  \right ) \left ( \sum_{\bm{k'}} (i \bm{k'}_{\bm{d}})^{*} A^{*}(\bm{k'})e^{-i \bm{k'}\bm{x}}  \right ) \\
&=\sum_{\bm{k},\bm{k'}}\bm{k}_{d}(\bm{k'}_{d})^{*} A(\bm{k})A^{*}(\bm{k'})d^{p} \sum_{\bm{j}}e^{i(\bm{k}-\bm{k'})\bm{x}} \\
&=\sum_{\bm{k},\bm{k'}}\bm{k}_{d}(\bm{k'}_{d})^{*} A(\bm{k})A^{*}(\bm{k'}) L^{p} \delta_{\bm{k},\bm{ k'}} \\
&= L^{p} \sum_{\bm{k}} |\bm{k}_{d} |^{2} |A(\bm{k})|^{2},
\label{CE_20}
\end{aligned}
\end{eqnarray}
where
\begin{eqnarray}
\bm{k}_{d} \equiv \left ( \frac{2\sin \left( \frac{k_{1}d}{2} \right) }{d} e^{i \frac{k_{1}d}{2}}, \dots, \frac{2\sin \left( \frac{k_{p}d}{2} \right) }{d} e^{i \frac{k_{p}d}{2}}  \right).
\end{eqnarray}
This expression for $\bm{k}_d$ can be derived from the definition of the discrete gradient Eq.~(\ref{DGra}) as
\begin{eqnarray}
e^{i[k_{1}(j_{1}+1)+k_{2}j_{2} \dots+k_{p}j_{p}]d}-e^{i[\bm{k}\bm{j}]d}=e^{i\bm{kx}}(e^{ik_{1}d}-1)=e^{i\bm{kx}}e^{i\frac{k_{1}d}{2}}2i\sin\frac{k_{1}d}{2} 
\end{eqnarray}
for the first component, and similarly for the other components. Note that $\bm{k}_{d} \to \bm{k} $ when $d \to 0$, and also that the difference between $\bm{k}$ and $\bm{k}_{d}$ is not negligible when $\bm{k}$ is lager than or comparable to $1/d$.

We now introduce an \textit{effective wavenumer} $k_{m}$:
\begin{eqnarray}
k^{2}_{m} \equiv \frac{\sum_{\bm{k}} |\bm{k}_{d}|^{2} |A(\bm{k})|^{2}}{\sum_{\bm{k} \ne 0}|A(\bm{k})|^{2}} \simeq  \frac{\sum_{\bm{k}} |\bm{k}_{d}|^{2} |\mu(\bm{k})|^{2}}{\sum_{\bm{k} \ne 0}|\mu(\bm{k})|^{2}},
\label{CE_31}
\end{eqnarray}
which characterizes the spatial arrangement of the active and inactive oscillators. The second expression follows from the approximation that the oscillators are binarized into two classes depending on the sign of $\mu$.
Plugging this definition into Eq.~(\ref{CE_20}) and using Parseval's identity, the summation can be approximated as
\begin{eqnarray}
\begin{aligned}
&d^{p} \sum_{\bm{j}} |\nabla A|^{2} = L^{p} k_{m}^{2}{\sum_{\bm{k} \ne 0}|A(\bm{k})|^{2}} \\
&=\frac{L^{p} k_{m}^{2}}{N^{p}}\sum_{\bm{j}}|A(\bm{j})-\overline{A}|^{2} \\
&=L^{p} k_{m}^{2}\left \{ \alpha_{S}[A_{S}-(\alpha_{S} A_{S} + \alpha_{H} A_{H})]^{2} + \alpha_{H}[A_{H}-(\alpha_{S} A_{S} + \alpha_{H} A_{H})]^{2}\right \} \\
&=L^{p}k_{m}^{2} \left \{ \alpha_{S} [\alpha_{H} A_{S} - \alpha_{H} A_{H}]^2 + \alpha_{H} [\alpha_{S}A_{H}-\alpha_{S}A_{S}]^{2}
\right \} \\
&=L^{p} k_{m}^{2} \alpha_{S} \alpha_{H} (\alpha_{S} + \alpha_{H})(A_{S}-A_{H})^{2} \\
&=L^{p} k_{m}^{2}\alpha_{S} \alpha_{H} (A_{S}-A_{H})^{2},
\end{aligned}
\label{CE_Gr}
\end{eqnarray}
where $\overline{A} = \alpha_S A_S + \alpha_H A_H$ denotes the direct-current component of $A$ in the DFT under the binarization assumption, i.e., the area average of $A$. Substituting Eq.~(\ref{CE_Am}) and Eq.~(\ref{CE_Gr}) into Eq.~(\ref{CE_F}), we obtain Eq.~(\ref{FE}).\\

\subsection{Stability condition}

Using the approximate GFE in Eq.~(\ref{FE}), we can derive the approximate conditions for the linear stability of the stationary quiescent state of the system.
The Hessian matrix of Eq.~(\ref{FE}) at the quiescent state $(A_{S},A_{H})=(0,0)$ is given by
\begin{eqnarray}
\begin{aligned}
M &\equiv \left [
\begin{array}{cc}
\frac{\partial^{2}F}{\partial A_{S} \partial A_{S}} & \frac{\partial^{2}F}{\partial A_{S} \partial A_{H}} \\
\frac{\partial^{2}F}{\partial A_{H} \partial A_{S}} & \frac{\partial^{2}F}{\partial A_{H} \partial A_{H}}
\end{array}
\right ]_{(A_{S},A_{H})=(0,0)} \\
&= \frac{1}{2} L^{p} \left[
\begin{array}{cc}
\alpha_{S}(\alpha_{H} r_{e}^{2} k_{m}^{2}-\mu_{S}) & -\alpha_{S} \alpha_{H} r_{e}^{2} k_{m}^{2} \\
-\alpha_{S} \alpha_{H} r_{e}^{2} k_{m}^{2} & \alpha_{H}(\alpha_{S} r_{e}^{2} k_{m}^{2}-\mu_{H})
\end{array}
\right ].
\end{aligned}
\end{eqnarray}
Thus, the quiescent state is linearly stable if and only if $\mbox{tr}(M)>0$ and $\mbox{det}(M)>0$, which results in the following two inequalities:
\begin{eqnarray}
\alpha_{S} \mu_{S}+\alpha_{H} \mu_{H}<0, \; \; \frac{\mu_{S} \mu_{H}}{\alpha_{S} \mu_{S}+\alpha_{H} \mu_{H}}<r_{e}^{2}k_{m}^{2}.
\label{CE_stable}
\end{eqnarray}

The first inequality indicates that the average value of $\mu$ should be negative for the linear stability of the system.  If the average value of $\mu$ is positive, the macroscopic oscillation cannot be suppressed by any $r_{e}$. 
This inequality is interpreted as the condition that the inactive group wins over the active group when $r_{e}$ becomes infinitely large.

The second inequality shows that larger diffusion constant $r_{e}$ is required to suppress the macroscopic oscillation when the effective wavelength of the arrangement of active and inactive oscillators is larger (i.e., when the effective wavenumber $k_{m}$ is smaller).
It is remarkable that both inequalities do not depend on $p$ and therefore they are valid in any dimensions.

\section{Application to the cortico-thalamic model}

To illustrate how our approximate theory can provide insights into realistic problems, we consider the cortico-thalamic model proposed by Kim and Robinson~\cite{kim2007compact, kim2009dynamics},
\begin{eqnarray}
\left[(t_e \frac{d}{dt}+1)^2-r_e^2 \Delta \right] \chi= (1+c_1) \chi
+c_2\tilde{\chi}+\epsilon\chi(t)^3,
\label{Compact}
\end{eqnarray}
where the field variable $\chi=\chi(j_{x}, j_{y}; t)$ represents a mean firing rate of the neurons within each local area of the cortex, $\tilde{\chi} \equiv \chi(j_{x}, j_{y}; t-t_{0})$ is a delayed value of $\chi$ (where $t_{0}$ is the time delay), $c_1$ parameterizes the strength of cortico-cortical activities, $c_2$ characterizes cortico-thalamic feedback, and $\epsilon$ controls the nonlinear term that takes into account the characteristics of neuronal firing.

The Kim-Robinson model is categorized into a neural \textit{field} model, which describes how a mesoscopic quantity characterizing neural activity evolves over both \textit{space} and time~\cite{pinotsis2014neural}. 
This model undergoes Hopf bifurcation and exhibits collective oscillations, which is interpreted as the onset of an epileptic seizure~\cite{kim2009dynamics}.
Though Eq.~(\ref{Compact}) has a simple expression, it is actually a system with time delay and exhibits a wide variety of oscillatory dynamics. It is known that the model can reproduce many of the typical oscillatory waveforms observed in real electroencephalograms.
By the center manifold reduction, the lattice complex Ginzburg-Landau model, Eq.~(\ref{CGL}), can be derived from Eq.~(\ref{Compact}) after
spatial discretization (see Ref.~\cite{yamaguchi2011reduction} for details). In the following, we consider Eq.~(\ref{CGL}) with the spatial arrangements of active and inactive oscillators assumed in Ref.~\cite{kim2009dynamics} and analyze the transition to macroscopic oscillations.

In Ref.~\cite{kim2009dynamics}, the authors analyzed focal epilepsy both numerically and analytically using the model Eq.~(\ref{Compact}) by assuming that the parameters $c_1$ and $c_2$ are in the unstable regime (i.e., the field variable $\chi$ undergoes oscillations) in the epileptic foci of the cortex, while $c_1$ and $c_2$ are in the stable regime ($\chi$ tends to vanish) in the remaining tissue.
Their study implied that areas  surrounding the epileptic foci play an essential role in the suppression of epileptic seizures. However, the analytical condition derived in Ref.~\cite{kim2009dynamics} was based on simplified assumptions and therefore not general enough.

In Ref.~\cite{kim2009dynamics}, it is assumed that the inactive region (inactive oscillators in our discrete model) never oscillates at all, there is only a single active area and, the oscillation decays at the edge of the boundary. These assumptions lead to the prediction that the active region can be suppressed whenever the diffusion constant $r_e$ is larger than a critical value. This prediction, however, is valid only in special cases.
Consider the limiting case where the proportion of the active area is close to $1$, where the activity of the active area is infinitely large, where the inactive area is vanishingly small, or where the inactivity (resistance to activation) of the inactive area is infinitely small. In these cases, it is obvious that the active region would not be suppressed no matter how large the diffusion constant $r_e$ is. As we show below, not the inactive group but the active group \textit{wins} under sufficiently strong coupling in these cases. 

The major difficulty in deriving a better analytical prediction is that Eq.~(\ref{Compact}) has a delay term, which complicates theoretical analysis.
This difficulty can be circumvented by reducing Eq.~(\ref{Compact}) with time delay to the lattice complex Ginzburg-Landau model Eq.~(\ref{CGL}) by the center manifold reduction near the bifurcation point, as we considered in Ref.~\cite{yamaguchi2011reduction}. 
In the reduced Eq.~(\ref{CGL}), epileptic foci and areas surrounding them correspond to the regions with $\mu>0$  and $\mu<0$, respectively.
See Ref.~\cite{yamaguchi2011reduction} for further details on Eq.~(\ref{Compact}) and its reduction to Eq.~(\ref{CGL}).
The other problem is that the stability condition for mixed populations of coupled active and inactive oscillators on a two-dimensional lattice is difficult to derive analytically. By using the approximate theory that we developed in the previous section, we can analyze the effect of spatial arrangement of the active and inactive oscillators, extending the result of Ref.~\cite{kim2009dynamics}.

In the following, we illustrate our theory by numerically integrating Eq.~(\ref{Compact}), which is spatially discretized on a 6 $\times$ 6 lattice, for several spatial arrangements of the oscillators.

\subsubsection{Investigated patterns}

\begin{figure}
\includegraphics[width=0.8\linewidth]{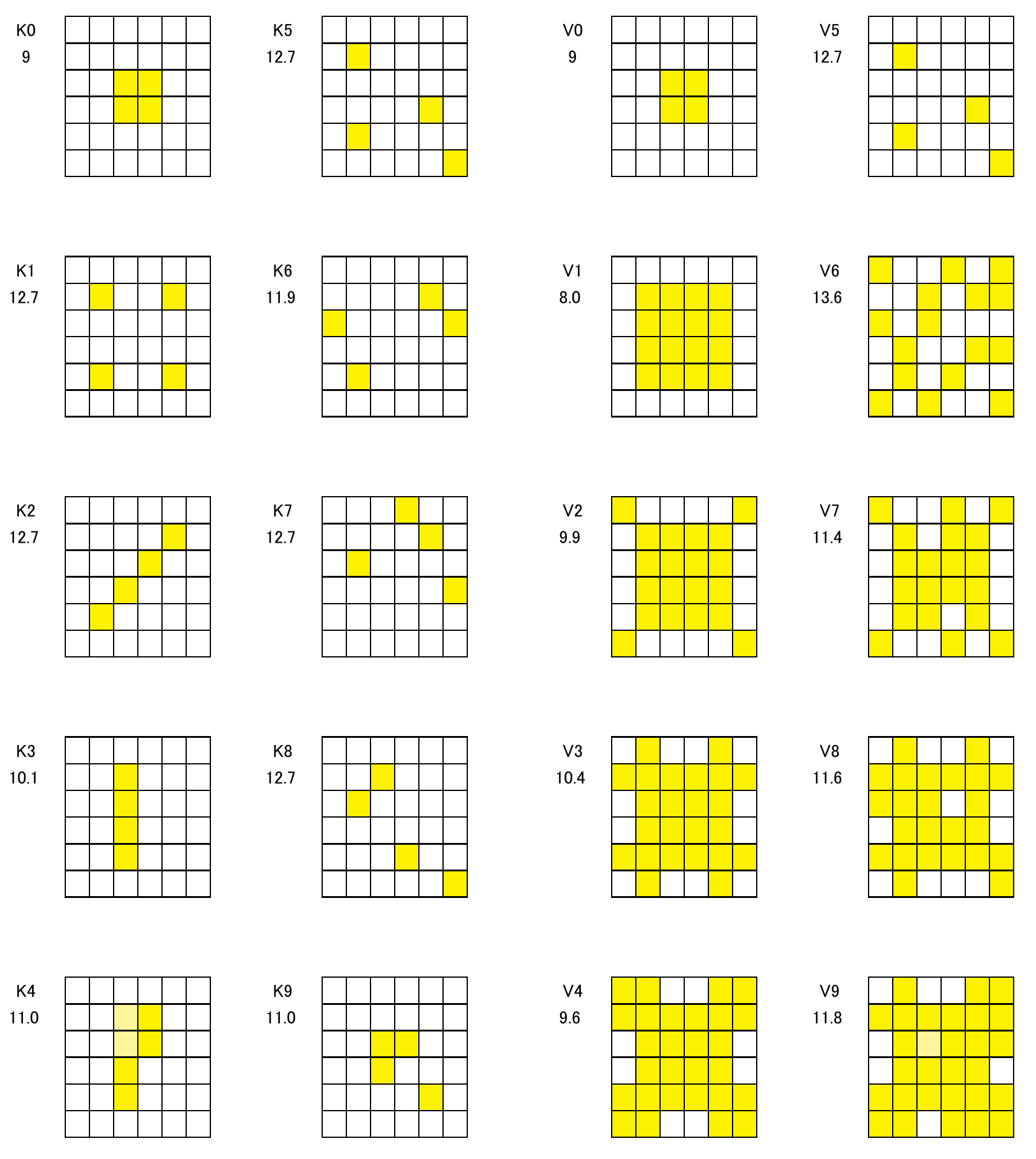}
\caption{(color online) Investigated spatial arrangements of the active and inactive oscillators. Yellow and white cells represent oscillators in group S and H, respectively. The pattern identifiers for the K-series and V-series are indicated in the left of the patterns, and the values below the identifiers represent the effective wavenumbers multiplied by the side length L, $k_mL$. \label{Patterns_PRE}}
\end{figure}

Figure~\ref{Patterns_PRE} shows the investigated arrangements of the active and inactive oscillators. The yellow cells represent active oscillators in group S, while the white cells represent inactive oscillators in group H.
We study the following two series of spatial arrangements: K-series (K0, K1, \dots, K9) and V-series (V0, V1, \dots, V9). The pattern identifier is indicated to the left of each pattern. The value below the identifier represents the effective wavenumber multiplied by the side length, $k_m L$, defined by Eq.~(\ref{CE_31}). 
In the K-series, $\alpha_{S}$ = 4/36 is fixed and all the patterns satisfy the first inequality of Eq.~(\ref{CE_stable}), while in the V-series, $\alpha_{S}$ varies from 4/36 to 28/36 and V3, V4, V8 and V9 do not satisfy the first inequality of Eq.~(\ref{CE_stable}). 

\subsubsection{Results for the K-series}

Using our approximate theory, we first reproduce the typical results presented in Ref.~\cite{kim2009dynamics}. Here, $L$ (= 60 cm) corresponds to the side length of a square of the cortex. There are four active oscillators at the center of the square as in the pattern K0 in Fig.1, which correspond to the focal epileptic area of the cortex. The parameter values used in our calculations are shown Table~\ref{CE_param_Kim}, which are the same as those used in Ref.~\cite{kim2009dynamics}.

In Ref.~\cite{kim2009dynamics}, it is shown that the macroscopic oscillation, which is considered to be a seizure, does not occur if $r_{e}$ is larger than about 2.5 cm in this case. 
This suggests that inhibition by the surrounding area of the epileptic focus plays an important role in the suppression of epileptic seizures.\\

\begin{table}[hbtp]
 \caption{Parameter values \label{CE_param_Kim}}
 \begin{center}
  \begin{tabular}{|c||c|c|c|c|c||c|c|}
   \hline
Parameter & $c_{1}$ & $c_{2}$ & $t_{e}$  & $t_{0}$ & $\epsilon$ & $\Omega$ & $\mu$ \\ \hline
Group S & -0.1 & -0.5 & 10msec  & 80msec & -0.1 & 24rad/sec & 0.116
\\ \hline
Group H & -0.4 & -0.5 & 10msec  & 80msec & -0.1 & 24rad/sec & -0.184
\\ \hline
  \end{tabular}
 \end{center}
\end{table}

\begin{figure}
\includegraphics[width=0.8\linewidth]{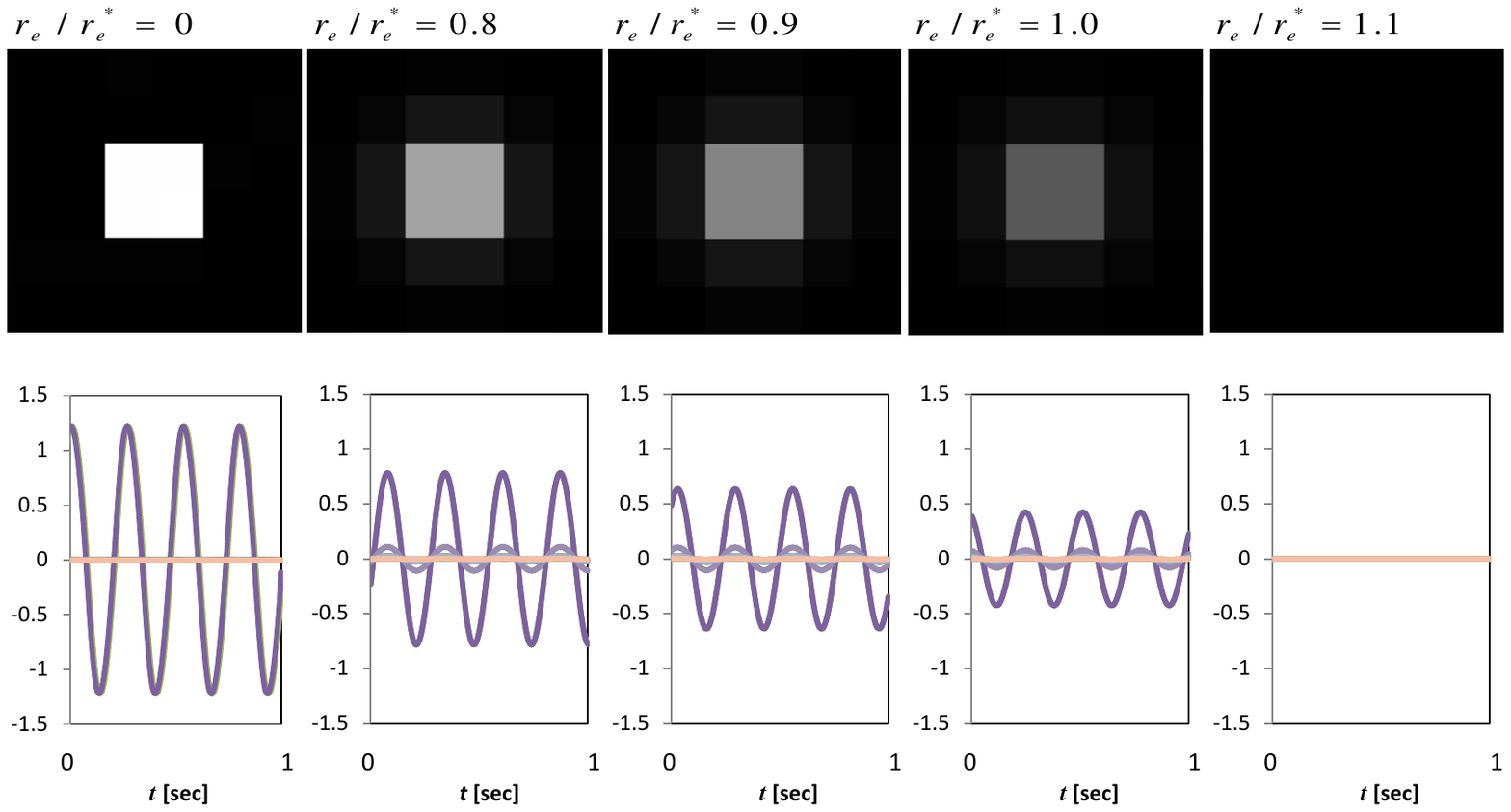}
\caption{(color online) Suppression of the macroscopic oscillation (occurrence of seizure) by the healthy region (inactive oscillators) surrounding the epileptic foci (active oscillators) at sufficiently large coupling strengths. In the top row, each figure shows the peak to peak amplitudes of the oscillators in 256-step grayscale (white corresponds to the maximum amplitude and black corresponds to zero amplitude). In the bottom row, each figure shows the waveforms of all oscillators sufficiently after the initial transient. \label{CE_K0}}
\end{figure}

Steady states of the system after a sufficiently long transient period are shown in Fig.~\ref{CE_K0} for several values of $r_{e}$. The top row shows peak to peak amplitude of the oscillators. 
 The waveforms of all oscillators are shown in the bottom row of Fig.~\ref{CE_K0}. When $r_{e}=0$, the oscillators are completely separated into two groups, i.e., oscillating group and quiescent group (the waveform of the quiescent group lies on the horizontal axis). As $r_{e}$ becomes larger, the inactive oscillators start to oscillate, while the active oscillators start to decrease their oscillation amplitudes, and all oscillators stop at the critical value of $r_{e}$. We evaluated the critical value of the diffusion constant $r_{e}^{*}$ from the stability condition given by Eq.~(\ref{CE_stable}) and obtained $r_{e}^{*}=$ 2.51 cm for the pattern K0. We can confirm that the oscillations actually disappear near this predicted point, in good agreement with Ref.~\cite{kim2009dynamics}.

\begin{figure}
\includegraphics[width=0.8\linewidth] {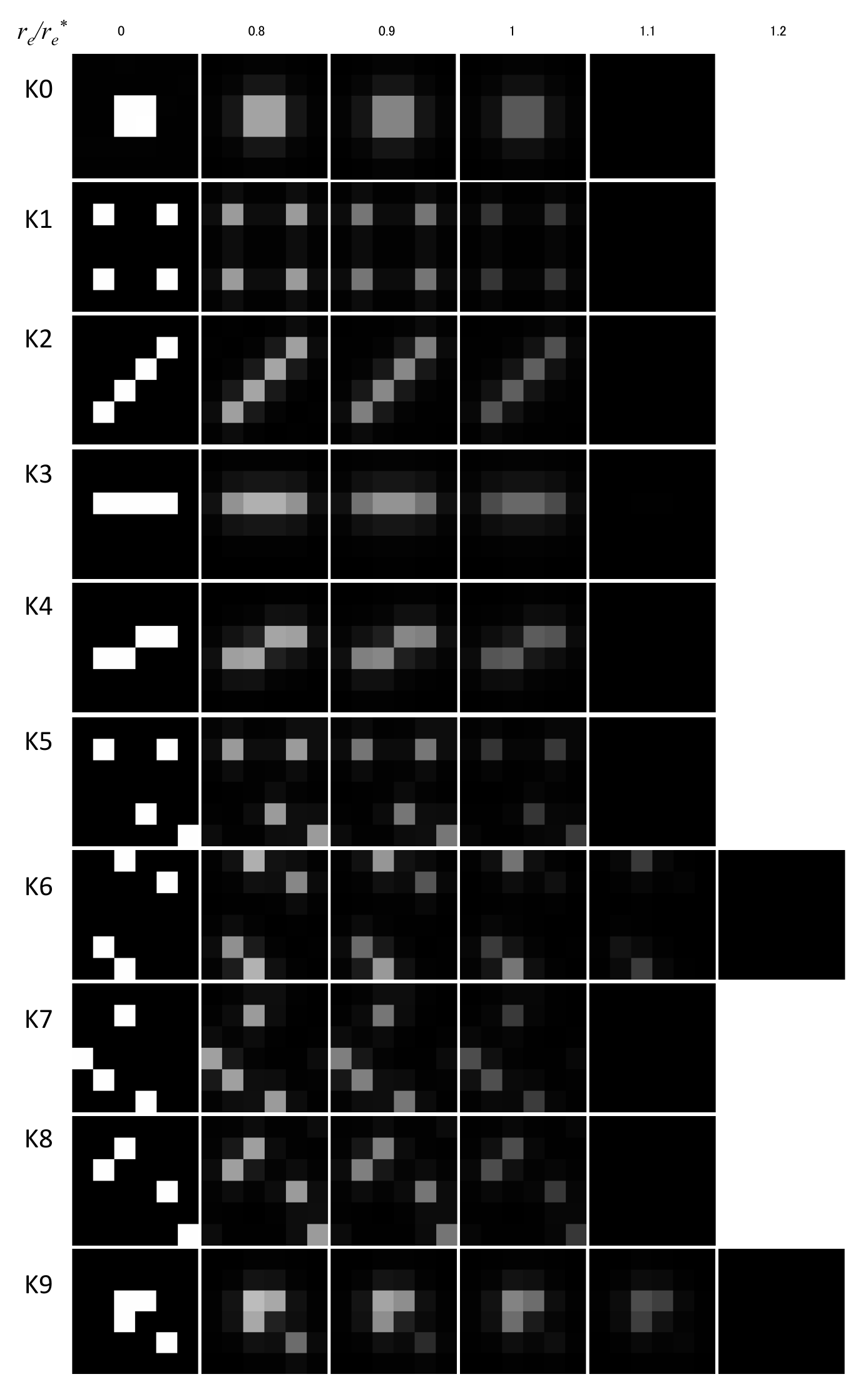}
\caption{(color online) Suppression of the macroscopic oscillation at sufficiently large coupling strength for the K-series patterns in Fig.~\ref{Patterns_PRE}. Peak to peak amplitudes of the oscillators are shown in grayscale for all patterns. \label{CE_K}}
\end{figure}

\begin{figure}
\includegraphics{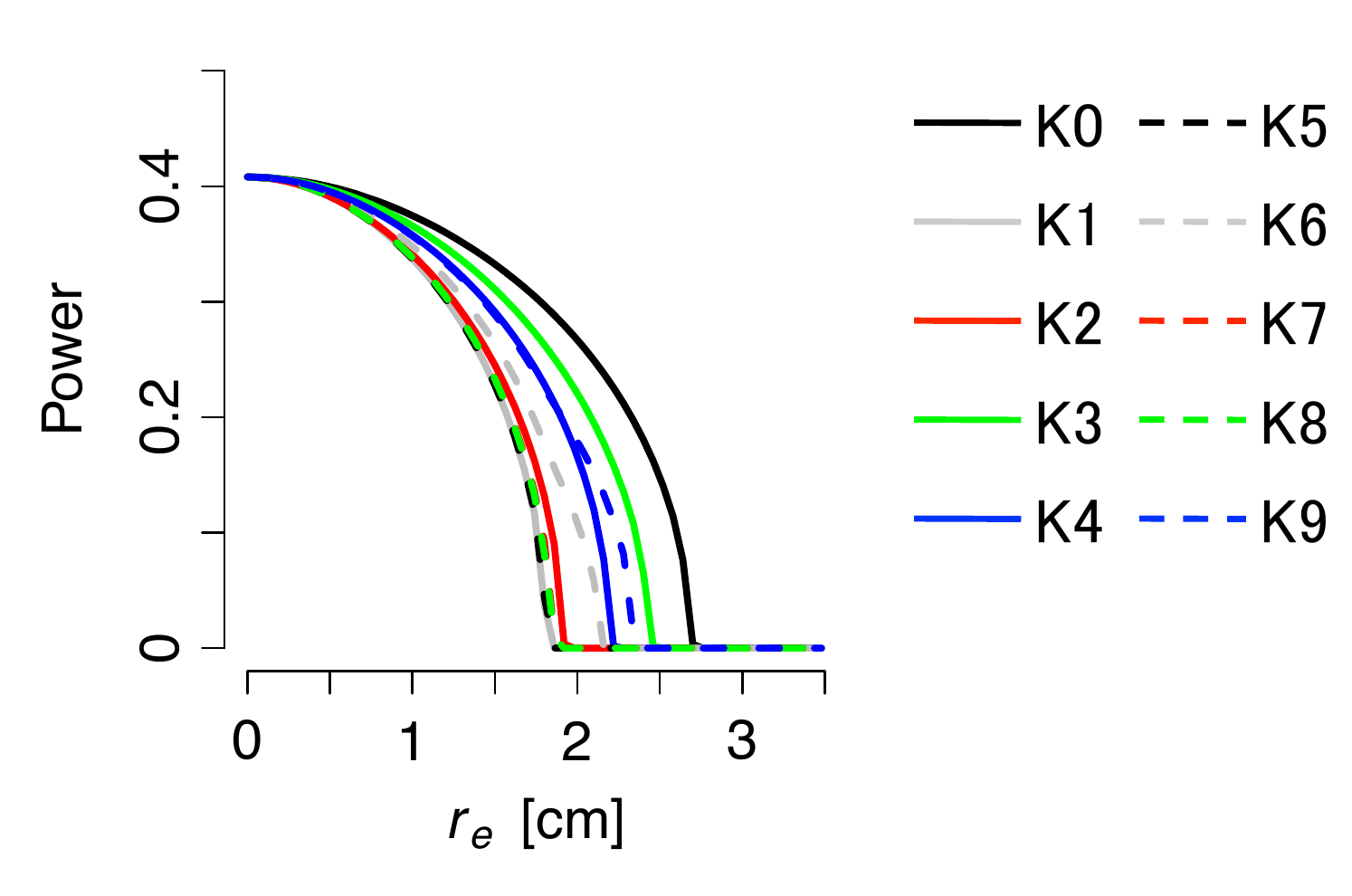}
\caption{(color online) Dependence of the total power of the macroscopic oscillation on the diffusion constant $r_{e}$ for the the K-series patterns in Fig.\ref{Patterns_PRE} \label{CE_K_power_abs}}
\end{figure}

Figure~\ref{CE_K} shows the results for all K-series patterns in Fig.\ref{Patterns_PRE} in grayscale, and Fig.~\ref{CE_K_power_abs} shows the power of the macroscopic oscillation vs. $r_{e}$, where the power is calculated as the areal and temporal mean square of the amplitudes of all oscillators. The oscillations disappear near the critical diffusion constants predicted by the present approximate theory. 

\subsubsection{Results for the V-series}

\begin{figure}
\includegraphics {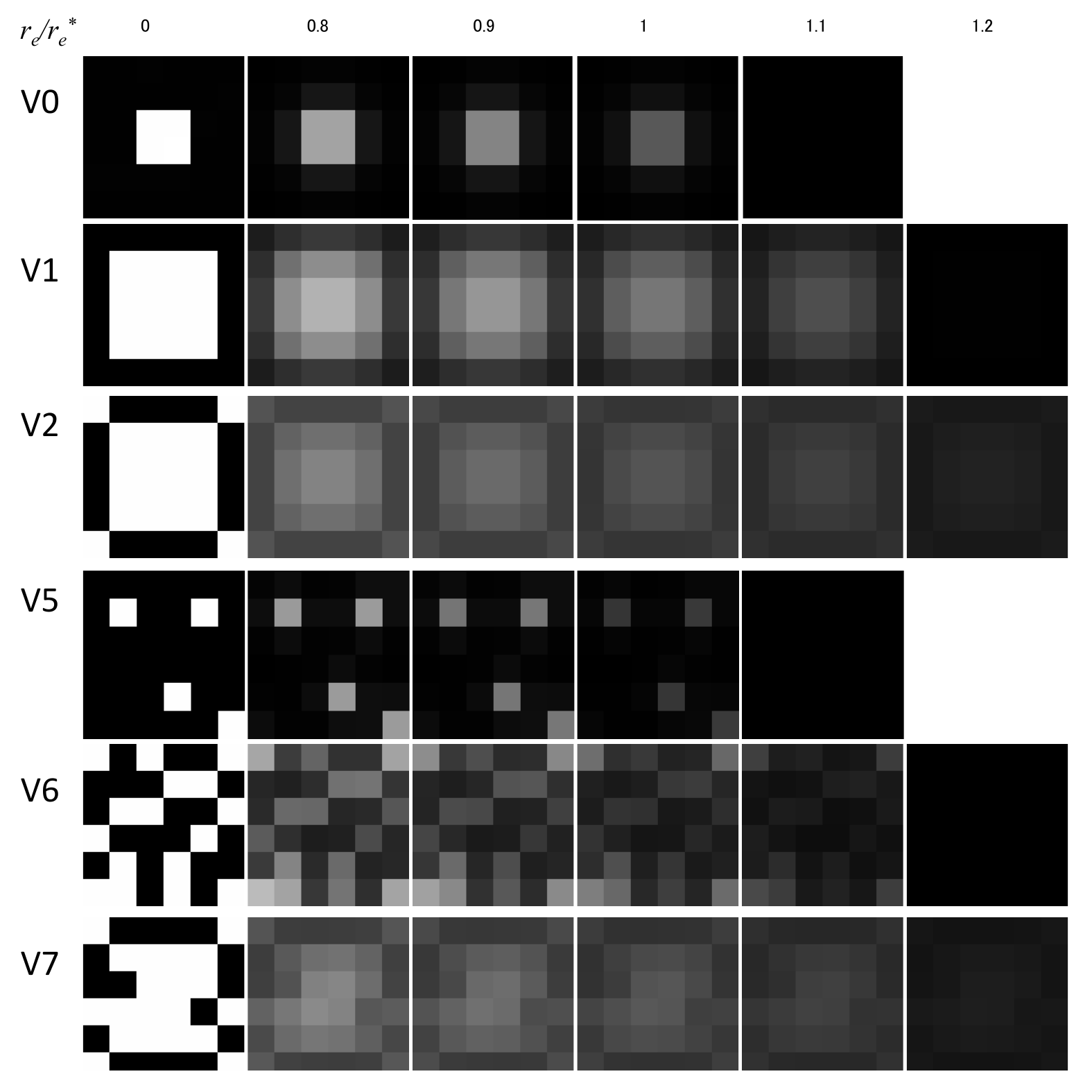}
\caption{(color online) Results for the V-series patterns in Fig.~\ref{Patterns_PRE}, where mean square amplitudes of the oscillators are shown in grayscale for all patterns in Fig.~\ref{Patterns_PRE} 
\label{CE_V}}
\end{figure}

\begin{figure}
\includegraphics [width=0.8\linewidth]{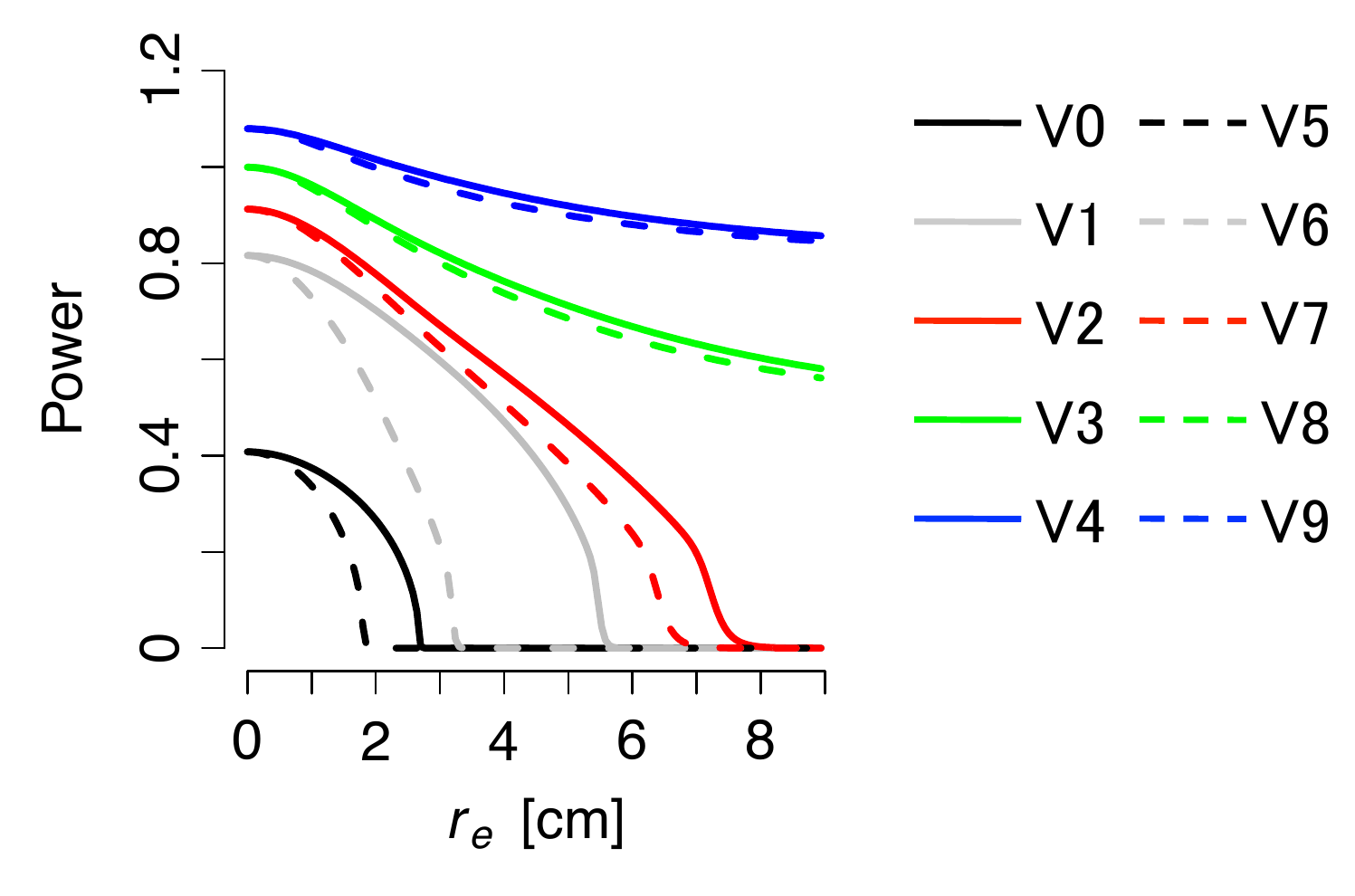}
\caption{(color online) Dependence of the total power of the macroscopic oscillation on the diffusion constant $r_{e}$ for the V-series patterns in Fig.~\ref{Patterns_PRE} \label{CE_V_p}}
\end{figure}

Figures~\ref{CE_V} and \ref{CE_V_p} show the results for the V-series patterns in Fig.~\ref{Patterns_PRE}.
In these cases, the oscillations should remain at any large $r_{e}$ in V3, V4, V8 and V9, because the first inequality of Eq.~(\ref{CE_stable}) is not satisfied in these patterns. The validity of the prediction can be seen in Fig.~\ref{CE_V_p}. As for the other patterns, we can again confirm that the oscillations disappear near the critical diffusion constants predicted by the second inequality of Eq.~(\ref{CE_stable}).   

The prediction errors are less than about 30\% of the true values through K-series and V-series. These errors are caused by the fact that the binarization assumption does not hold exactly, especially when the characteristic length scale of the patterns is large. But the error is still not fatal and does not undermine the insight; we can still predict whether the suppression of the macroscopic oscillation occurs or not by the approximate theory.

\section{DISCUSSION AND CONCLUSION}

We have derived an approximate condition for the suppression of macroscopic oscillations  in mixed populations of coupled active and inactive oscillators using the two-group approximation for the lattice complex Ginzburg-Landau model, which can be considered an aging transition due to local diffusive coupling. As illustrated by numerical simulations, the approximate condition, Eq.~(\ref{CE_stable}), explains the transitions qualitatively well.
Although quantitative accuracy is not very well and further improvement in the approximation is desirable, we believe that the theory
developed in this paper provides an important step toward understanding of the aging transition in mixed populations coupled oscillators.

In the present study, we focused our attention on the generalized free energy (GFE) and approximated it to analyze the transition in a simple way.
Although the approximate GFE does not have the same information as the original dynamical equation in the sense that we cannot derive the original equation from the approximate GFE, it contains essential information for predicting the stability and helps us derive the stability condition.
Note that, unlike the real Ginzburg-Landau equation, several conditions must be satisfied for the complex Ginzburg-Landau equation to possess a GFE function.

In our analysis, Fourier transform to the wavenumber space was used to extract the effective wave number $k_{m}$ of the spatial arrangement of the mixed oscillator population, which played an important role in determining the stability of the quiescent state in addition to the proportions of the oscillators ($\alpha_{S}, \alpha_{H}$) and the bifurcation parameters  ($\mu_{S}, \mu_{H}$). 
The following proposition is useful in understanding the meaning of $k_m$ and in calculating its actual value: the binding number $n_b$ between group S and group H is related to the effective wavenumber $k_{m}$ as $n_{b}=N^{p} \alpha_{S}\alpha_{H} (k_{m}d)^{2}$. We can prove this proposition by noting that $|\nabla \mu|^{2}$ takes a non-zero value $|(A_S-A_H)/d|^{2}$ only at the boundaries between the group S and group H under our binarization assumption. Plugging $|\nabla \mu|^{2}=|(A_S-A_H)/d|^{2}$ at the boundaries and otherwise $|\nabla \mu|^{2}=0$ into Eq.~(\ref{CE_Gr}) leads
\begin{equation}
d^{p-2} n_{b}|(A_{S}-A_{H})|^{2} = L^{p} k_{m}^{2}\alpha_{S}\alpha_{H}|(A_{S}-A_{H})|^{2},
\end{equation}
which gives
$n_{b}=N^{p} \alpha_{S}\alpha_{H} (k_{m}d)^{2}$, using $L=Nd$.\\

Finally, though we illustrated the theory for a model of epileptic seizures, we stress that our result is generally applicable to coupled-oscillator models near the Hopf bifurcation, because the complex Ginzburg-Landau equation is a normal form of coupled oscillators near the Hopf bifurcation derived by the center manifold reduction method from the original model ~\cite{yamaguchi2011reduction, yang2017critical}.

\begin{acknowledgements}
The authors would like to thank Yoshiharu Yamamoto, Toru Nakamura, Fumiharu Togo, Akifumi Kishi and Jerome Foo for useful discussion. This work is partly supported by JSPS KAKENHI Grant number 15K01499 and 18K17887 to IY.
\end{acknowledgements}

\bibliography{reference.bib}

\newpage


\end{document}